\documentclass[a4paper,10pt]{article}

\usepackage{graphicx}
\usepackage{tikz}
\usepackage{amsmath}
\usepackage{amssymb}

\usepackage{dcolumn}
  \newcolumntype{d}[1]{D{.}{.}{#1}}

\makeatletter
\def\Ddots{\mathinner{\mkern1mu\raise\p@
\vbox{\kern7\p@\hbox{.}}\mkern2mu
\raise4\p@\hbox{.}\mkern2mu\raise7\p@\hbox{.}\mkern1mu}}
\makeatother

\newcommand{\Var}{\operatorname{Var}}
\newcommand{\Corr}{\operatorname{Corr}}
\newcommand{\SCR}{\operatorname{SCR}}
\newcommand{\prob}[1]{\operatorname{P}\left( #1 \right)}
\newcommand{\M}{\mathcal{M}}
\newcommand{\E}{\mathbb{E}}

\newcommand{\scaleFactorOne}{0.82}

\newcommand{\tsub}[1]{_{\text{#1}}}
\newcommand{\tsup}[1]{^{\text{#1}}}

\title{Signs of dependence and heavy tails in non-life insurance data}
\author{Jonas Alm\\Chalmers University of Technology, Gothenburg, Sweden}
\date{\today}

\begin{document}

\maketitle

\begin{abstract}
\noindent In this paper we study data from the yearly reports the four major Swedish non-life insurers have sent to the Swedish Financial Supervisory Authority (FSA). We aim at finding marginal distributions of, and dependence between, losses on the five largest lines of business (LoBs) in order to create models for Solvency Capital Requirement (SCR) calculation. We try to use data in an optimal way by sensibly defining an accounting year loss in terms of actuarial liability predictions, and by pooling observations from several companies when possible to decrease the uncertainty about the underlying distributions and their parameters. We find that dependence between LoBs is weaker in our data than what is assumed in the Solvency II standard formula. We also find dependence between companies that may affect financial stability, and must be taken into account when estimating loss distribution parameters. Moreover, we discuss under what circumstances an insurer is better (or worse) off using an internal model for SCR calculation instead of the standard formula.
\end{abstract}

\section{Introduction}

In recent years much research has been focused on new ways to model dependence of insurance risks (for example \cite{DiersEtAl12}, \cite{LeeLin12}, \cite{MerzEtAl13} and \cite{SalzmannWuthrich12}). Some interesting work has also been done on what to do in the case when data for dependence estimation is scarce (see \cite{BernardEtAl14} and \cite{EmbrechtsEtAl13}). Less research has been aimed at what marginal distributions to use when modeling insurance losses; in the Solvency II standard formula for Solvency Capital Requirement (SCR) calculation one simply assumes the lognormal distribution and then calibrates the parameters to get an overall SCR that corresponds to the 99.5\% Value-at-Risk (VaR) of the insurer's basic own funds over a one-year period, ie, the 0.995 quantile of the insurer's one-year loss distribution (see \cite{Eiopa14b} for details). In this context, the one-year loss of an insurer is the change in the insurer's net liability value over a one-year horizon. If we only consider risks on the liability side of the balance sheet the one-year loss is the change in prediction of liabilities from one year to the next.

As most practitioners know, the limiting factor when trying to come up with reasonable capital requirements for an insurance company is the lack of useful data. The author's experience is that the best one can hope for is about 20 observations of yearly losses for each line of business (LoB) of an insurance company, and that these observations often must be adjusted in order to be viewed as identically distributed. If one could pool data from several companies the number of observations would increase dramatically, but since the companies to a large extent are exposed to the same risks dependence must be examined carefully before pooling data. Moreover, testing whether or not the samples from the different companies can be considered drawn from the same distribution is crucial. For example, if we pool data from samples with different variances, the pooled sample may show signs of heavy-tailedness that is not present in the individual samples.

In this paper we study data from the yearly reports the four major Swedish non-life insurers (\emph{Folksam}, \emph{If}, \emph{L\"{a}nsf\"{o}rs\"{a}kringar (LF)} and \emph{Trygg-Hansa}) have sent to the Swedish Financial Supervisory Authority (FSA) (\emph{Finansinspektionen} in Swedish). We aim at finding marginal distributions of, and dependence between, the five LoBs \emph{Illness and Accident}, \emph{Home}, \emph{Business Liability and Property}, \emph{Motor Liability} and \emph{Motor Other} to create models for SCR calculation solely based on the FSA data. We investigate if the dependence structure we see in data agrees with the dependence structure in the Solvency II standard formula. Moreover, we try to find signs of dependence between the companies. Too much dependence between companies is bad for financial stability. Also, if the dependence between companies is not too strong we may pool data from several companies to get less uncertainty in the loss distribution parameter estimation.

This is the outline of the paper: In Section~\ref{sec:notation} we introduce the notation, set up liability cash flow predictions and define an accounting year loss in terms of changes in these predictions. We explain how data from the FSA reports are transformed into losses in Section~\ref{sec:data}. In Section~\ref{sec:dataAnalysis} we analyze tail-heaviness of the insurance loss data, and investigate the dependence between LoBs and companies. We give suggestions of how to model marginal distributions and dependence to calculate SCR in Section~\ref{sec:modeling}, and in Section~\ref{sec:standardFormula} we compare our models to the standard formula in Solvency II. Section~\ref{sec:discussion} contains a concluding discussion.

\section{Notation} \label{sec:notation}

In this section we introduce the notation that will be used throughout the paper. In particular, we define an accounting year loss in terms of changes in liability predictions.

Consider a fixed company and a fixed LoB, and denote the cumulative claim payments for accident year $i$ and development year $j$ by $C_{i,j}$. This is the total amount paid by the end of accounting year $i+j$ for claims arising from accidents incurred during year $i$. The actuarial ultimo prediction at time $n$ (ie, the end of accounting year $n$) of cumulative claim payments for accident year $i$ is denoted by $\hat{C}_{i,\text{ultimo}}^{(n)}$, and the earned premium for accident year $i$ is denoted by $V_i$.

Now, consider accounting year $n+1$. In the beginning of the accounting year, the (undiscounted) value of the cash flow arising from already incurred claims is a summation of predicted payments over the last $k-1$ accident years,
\begin{equation}
R_0 := \sum_{i=n-k+2}^{n} \left( \hat{C}_{i,\text{ultimo}}^{(n)} - C_{i,n-i} \right).
\label{eq:R0}
\end{equation}
(We use the letter $R$ since the change in valuation of this cash flow is related to the reserve risk.) To value the cash flow arising from not yet incurred claims we define the loss ratio
\begin{equation}
\hat{\ell}_{n+1}^{(n)} := \left. \sum_{i=n-m+1}^{n} \hat{C}_{i,\text{ultimo}}^{(n)} \middle/ \sum_{i=n-m+1}^{n} V_i, \right.
\label{eq:LossRatio}
\end{equation}
where $m$ is the number of years used for estimation. Given a choice of $m$, the value of the cash flow arising from claims incurring in year $n+1$ (not yet incurred claims) is
\begin{equation}
P_0 :=  V_{n+1} \hat{\ell}_{n+1}^{(n)}.
\label{eq:P0}
\end{equation}
(We use the letter $P$ since the change in valuation of this cash flow is related to the premium risk.) Note that we assume $V_{n+1}$ known in the beginning of accounting year $n+1$ even though, in reality, it is not completely known before the year ends. This is not a very strong assumption since the prediction error of the earned premium in general is very small compared to the errors in the payment predictions. Moreover, if the earned premium is lower (higher) than what we predicted in the beginning of the accounting year then the loss will also be lower (higher), so the error does not matter substantially.

The prediction of the total outstanding liability cash flow is now
\begin{equation}
Y_0 := R_0 + P_0.
\label{eq:Y0}
\end{equation}
In the end of the accounting year, we have new information and hence new valuations of the above cash flows. These valuations are
\begin{align}
& R_1 := \sum_{i=n-k+2}^{n} \left( \hat{C}_{i,\text{ultimo}}^{(n+1)} - C_{i,n-i} \right), \label{eq:R1} \\
& P_1 :=  \hat{C}_{n+1,\text{ultimo}}^{(n+1)}, \label{eq:P1} \\
& Y_1 := R_1 + P_1. \label{eq:Y1}
\end{align}
We define a loss as a positive change in the valuation of outstanding liabilities. The normalized loss for accounting year $n+1$ is given by
\begin{equation}
U := \frac{Y_1 - Y_0}{Y_0}.
\label{eq:U}
\end{equation}

\section{Data} \label{sec:data}

In this section we explain which data in the FSA reports we use and how we transform these data into losses on the different LoBs of each company.

The data comes from the yearly reports Folksam, If, LF and Trygg-Hansa have sent to the FSA. The five LoBs we consider are: Illness and Accident (IA), Home (H), Business Liability and Property (BLP), Motor Liability (ML) and Motor Other (MO). The reports contain, for each LoB, today's cumulative claim payments and ultimo predictions of the cumulative claim payments (in nominal values) for the last $k$ accident years. For the LoBs Home and Motor Other $k=3$, for Illness and Accident and Business Liability and Property $k=10$, and for Motor Liability $k=15$. The reports also contain values of the (nominal) earned premiums for the three latest accident years.

The FSA data assumed known at time $n$ are shown in Table~\ref{tab:reportdata}. As mentioned in the previous section, we consider $V_{n+1}$ known already at time $n$ even though it is not completely known until time $n+1$.
\begin{table}[tb]
\centering
\scalebox{\scaleFactorOne}{
\begin{tabular}{|l|llllll|}
\hline
Earned premium & \multicolumn{6}{|c|}{Cumulative payments and ultimo predictions} \\
\hline
 & & & & $C_{n-k+1,k-1}$ & & $\hat{C}_{n-k+1,ultimo}^{(n)}$ \\
$V_{n-2}$ & & & $\Ddots$ & & & \vdots \\
$V_{n-1}$ & & $C_{n-1,1}$ & & & & $\hat{C}_{n-1,ultimo}^{(n)}$ \\
$V_{n}$ & $C_{n,0}$ & & & & & $\hat{C}_{n,ultimo}^{(n)}$ \\
\hline
$V_{n+1}$ & & & & & & \\
\hline
\end{tabular}
}
\caption{Data considered known at time $n$ (ie, the end of accounting year $n$ or beginning of accounting year $n+1$).}
\label{tab:reportdata}
\end{table}
For each accounting year, we use the data known in the beginning of the year and make a prediction ($Y_0$) of the outstanding liability cash flow using \eqref{eq:R0}--\eqref{eq:Y0}, with $m=3$ in \eqref{eq:LossRatio}. Then, we move forward in time until the end of the year when the next report is available and make a new prediction ($Y_1$) of the same cash flow using \eqref{eq:R1}--\eqref{eq:Y1}. The normalized one-year loss ($U$) is now given by \eqref{eq:U}.

The ultimo predictions in \eqref{eq:R0}, \eqref{eq:R1} and \eqref{eq:P1} were made by actuaries at the different companies, but the methods used are not stated in the reports. The only intervention by the author to the normalized loss construction above is the choice $m=3$ in \eqref{eq:LossRatio}. We make this choice since we only have three years of premium data in the first available FSA report. We could use a longer period for later accounting years, and this may give better predictions if the loss ratio is stationary. However, it is reasonable to believe that changes in pricing procedures make the loss ratio non-stationary. If we use $m=1$ or $m=2$ we get normalized losses very similar to the case $m=3$.

\begin{figure}[tb]
\centering
\includegraphics[width=\textwidth]{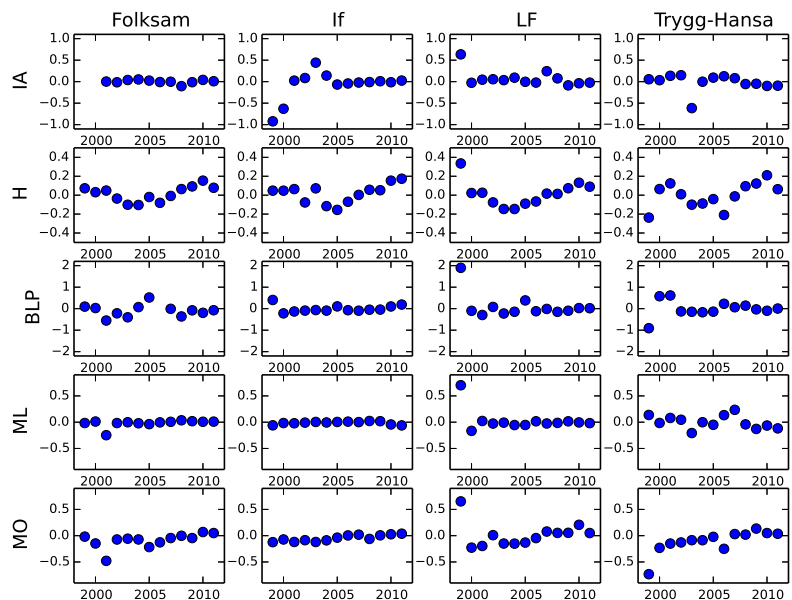}
\caption{Time series plots of the normalized losses ($U$) for the accounting years 1999 to 2011. Three of Folksam's losses are not visible in the plots, two in IA and one in BLP.}
\label{fig:LossOverview}
\end{figure}

We construct normalized losses ($U$) for the accounting years 1999 to 2011 for all five LoBs in the four companies. The reports considered in this paper were introduced in 1998, so we cannot go further back in time. As seen in Figure~\ref{fig:LossOverview}, the data quality of the first one to three accounting years is questionable. This may be due to misreporting, Folksam's first two normalized losses in the LoB Illness and Accident are for example not even visible in the plot. Folksam also has one large normalized loss in the LoB Business Liability and Property which is not visible in the plot. However, since Folksam sold most of this LoB to Trygg-Hansa in 2001, the absolute loss is small for a company of Folksam's size.

For the LoBs Illness and Accident and Motor Liability we consistently have $R_i >> P_i$ ($i=0,1$), so revaluation and payments of claims incurred \emph{before} the considered accounting year is the main component of the loss for these LoBs. For the LoBs Home and Motor Other we consistently have $P_i >> R_i$, so re-valuation and payment of claims incurred \emph{during} the considered accounting year is the main component of the loss for these LoBs. For the LoB Business Liability and Property $R_i$ and $P_i$ are similar in size.

\section{Data analysis} \label{sec:dataAnalysis}

In this section we study marginal distributions of the different LoBs, dependencies between companies for a specified LoB, and dependencies between LoBs within companies.

In Figure~\ref{fig:LossOverview} we see that the normalized losses are centered around zero for all LoBs and companies so there is no obvious bias in the liability predictions. For the LoB Motor Liability, the variance seems to be higher for Trygg-Hansa than for the other three companies. The author is not aware of any insurance events that have affected Trygg-Hansa more than the other companies that could explain this fact. Moreover, there seem to be positive trends in the losses for the LoBs Home and Motor Other, at least from 2004. The explanation may be that the winters 2009/10, 2010/11 and 2011/12 were colder than the winters in the first years of the 21st century. Another hypothesis is that competition has become fiercer leading to lower premiums so that a predictor based on historical loss ratios underestimates future losses.

\begin{figure}[tb]
\centering
\includegraphics[width=\textwidth]{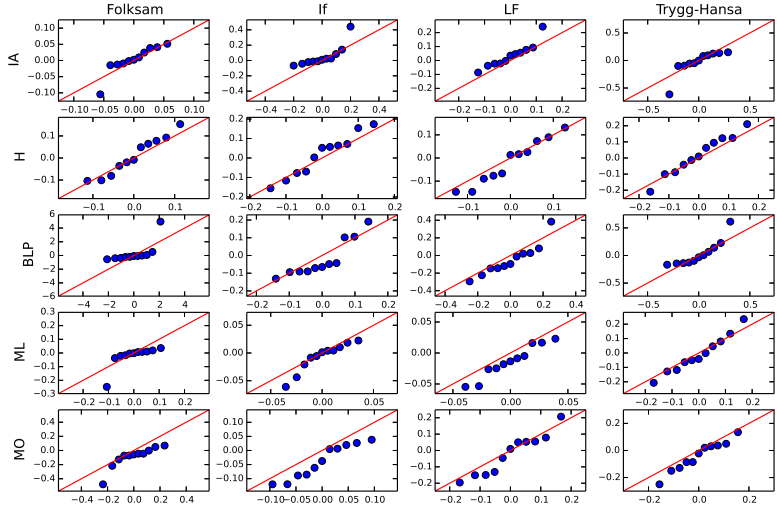}
\caption{Normal QQ plots based on normalized losses ($U$) for the accounting years 2001 to 2011 assuming zero mean for each LoB in each company.}
\label{fig:NormalQQplots}
\end{figure}

In the analysis that follows we exclude the losses for the first two accounting years for all LoBs and companies due to questionable data quality. The QQ plots in Figure~\ref{fig:NormalQQplots} indicate that the normal distribution is a reasonable choice for the marginal distributions of losses on the LoBs Home, Motor Liability and Motor Other. Folksam has two outliers, one in the left tail of Motor Liability and one in the left tail of Motor Other. These outliers are both from accounting year 2001 which is the third year in the period, and it is unclear whether these outliers exist due to misreporting or some insurance events. In the LoBs Illness and Accident and Business Liability and Property there are signs of heavy tails in the data.

As mentioned in the previous section, Folksam's Business Liability and Property LoB is small both compared to Folksam's other LoBs and to the other three companies' BLP LoBs. There is also one extreme outlier (with value 4.94) among Folksam's normalized BLP losses leading to a much larger sample standard deviation for Folksam than for the other three companies (see Table~\ref{tab:SampleStandardDeviations}). Due to Folksam's different size and data it does not make sense to include Folksam when pooling BLP data trying to improve distributional assumptions and parameter estimates. We also leave Folksam's BLP data out when studying dependence between companies and LoBs.

\begin{table}[tb]
\centering
\scalebox{\scaleFactorOne}{
\begin{tabular}{ld{1.3}d{1.3}d{1.3}d{1.3}}
\hline
& \multicolumn{1}{c}{Folksam} & \multicolumn{1}{c}{If} &  \multicolumn{1}{c}{LF} & \multicolumn{1}{c}{Trygg-Hansa} \\
\hline
IA & 0.040 & 0.15 & 0.090 & 0.21 \\
H & 0.082 & 0.10 & 0.092 & 0.12 \\
BLP & 1.5 & 0.10 & 0.18 & 0.22 \\
ML & 0.077 & 0.026 & 0.028 & 0.12 \\
MO & 0.17 & 0.069 & 0.12 & 0.11 \\
\hline
\end{tabular}
}
\caption{Sample standard deviations of the normalized losses for the accounting years 2001 to 2011 assuming zero mean for each LoB in each company.}
\label{tab:SampleStandardDeviations}
\end{table}

To check whether or not pooling of data makes sense, we conduct Levene's tests for equality of variances (see \cite{Levene60} and \cite{BrownForsythe74} for details) for each LoB. The test statistic $W$ is based on absolute deviations from the medians. Let $u_{ij}$ be the $j$th observed normalized loss and $m_i$ be the median of the observed normalized losses for company $i$. If $z_{ij} = | u_{ij} - m_i | $, then
\[
W = \frac{g (n-1) }{ g - 1}  \frac{ n \sum_{i=1}^{g} (\bar{z}_{i \cdot} - \bar{z}_{\cdot \cdot})^2 } { \sum_{i=1}^{g} \sum_{j=1}^{n} (z_{ij} - \bar{z}_{i \cdot})^2 },
\]
where $g$ is the number of companies, $n$ is the number of observations for each company, $\bar{z}_{i \cdot} = \sum_{j=1}^{n} z_{ij} / n$ is the company mean and $\bar{z}_{\cdot \cdot} = \sum_{i=1}^{g} \bar{z}_{i \cdot} / g$ is the overall mean. Under the assumption of equal variances across all samples, $W$ is $F$-distributed with $g-1$ and $g(n-1)$ degrees of freedom.

Since Trygg-Hansa have a higher Motor Liability sample variance than the other companies, we conduct Levene's tests both with and without Trygg-Hansa in the sample set for this LoB. The test statistics shown in Table~\ref{tab:TestStatistic} clearly indicate that the Motor Liability variance for Trygg-Hansa differs from the variances of the other companies. For the other LoBs it seems reasonable to conclude that the samples for the different companies have similar variances. Note that Folksam is not included in the BLP sample set.

\begin{table}[tb]
\centering
\scalebox{\scaleFactorOne}{
\begin{tabular}{ld{1.2}d{1.4}}
\hline
Line of Business & \multicolumn{1}{c}{Test statistic ($\hat{W}$)} & \multicolumn{1}{c}{$p$-value} \\
\hline
Illness and Accident & 1.93 & 0.14 \\
Home & 0.36 & 0.78 \\
Business Liability and Property (without Folksam) & 1.14  & 0.33 \\
Motor Liability & 4.71 & 0.0066 \\
Motor Liability (without Trygg-Hansa) & 0.64 & 0.54 \\
Motor Other & 0.82 & 0.49 \\
\hline
\end{tabular}
}
\caption{Observed values of Levene's test statistic $W$ and the corresponding $p$-values.}
\label{tab:TestStatistic}
\end{table}

Rank correlations, in terms of Spearman's $\rho$, between losses of different companies for each LoB are shown in Table~\ref{tab:RankCorrelationsBetweenCompanies}, and rank correlations between losses of different LoBs within the companies are shown in Table~\ref{tab:RankCorrelationsBetweenLoBs}. We choose Spearman's $\rho$ as a measure of association since it is both robust against non-normality and close to the linear correlation coefficient in the bivariate normal case,
\[
\rho\tsub{Spearman's} = \frac{6}{\pi} \arcsin{\frac{\rho\tsub{linear}}{2}}.
\]
For two independent samples with 11 observations each, the probability of getting a Spearman's $\rho$ with absolute value larger than 0.62 is 5\%. Thus, it is unlikely that the losses of different companies are independent for the LoBs Home and Motor Other, see Table~\ref{tab:RankCorrelationsBetweenCompanies}. For the other LoBs it seems reasonable to assume independence between companies. Based on the scattered rank correlations between LoBs within the companies in Table~\ref{tab:RankCorrelationsBetweenLoBs}, it seems reasonable to assume independence between LoBs except between Home and Motor Other. This is the only pair that has correlations of the same sign for all companies.

\begin{table}[tb]
\centering
\scalebox{\scaleFactorOne}{
\begin{tabular}{ld{2.2}d{2.2}d{2.2}d{2.2}d{2.2}}
\hline
 & \multicolumn{1}{c}{IA} & \multicolumn{1}{c}{H} & \multicolumn{1}{c}{BLP} & \multicolumn{1}{c}{ML} & \multicolumn{1}{c}{MO} \\
\hline
(Folksam, If)  & 0.21 & 0.58 & \multicolumn{1}{c}{-} & 0.30 & 0.63 \\
(Folksam, LF) & -0.09 & 0.93 & \multicolumn{1}{c}{-} & 0.17 & 0.78 \\
(Folksam, Trygg-Hansa) & -0.40 & 0.87 & \multicolumn{1}{c}{-} & -0.42 & 0.79 \\
(If, LF) & 0.18 & 0.67 & 0.51 & 0.14 & 0.79 \\
(If, Trygg-Hansa) & -0.25 & 0.51 & -0.23 & -0.06 & 0.67 \\
(LF, Trygg-Hansa) & 0.32 & 0.79 &  -0.31 & -0.04 & 0.76 \\
\hline
\end{tabular}
}
\caption{Rank correlations between companies for each LoB in terms of Spearman's $\rho$.}
\label{tab:RankCorrelationsBetweenCompanies}
\end{table}

\begin{table}[tb]
\centering
\scalebox{\scaleFactorOne}{
\begin{tabular}{ld{2.2}d{2.2}d{2.2}d{2.2}}
\hline
& \multicolumn{1}{c}{Folksam} & \multicolumn{1}{c}{If} &  \multicolumn{1}{c}{LF} & \multicolumn{1}{c}{Trygg-Hansa} \\
\hline
(IA, H) & -0.22 & 0.26 & -0.54 & -0.13 \\
(IA, BLP) & \multicolumn{1}{c}{-} & -0.22 & -0.27 & 0.30 \\
(IA, ML) & -0.36 & -0.25 & -0.50 & 0.76 \\
(IA, MO) & 0.05 & -0.52 & -0.26 & -0.65 \\
(H, BLP) & \multicolumn{1}{c}{-} & 0.29 & 0.16 & 0.34 \\
(H, ML) & 0.60 & -0.34 & 0.53 & -0.17 \\
(H, MO) & 0.58 & 0.27 & 0.65 & 0.52 \\
(BLP, ML) & \multicolumn{1}{c}{-} & -0.01 & -0.48 & 0.49 \\
(BLP, MO) & \multicolumn{1}{c}{-} & 0.55 & 0.46 & -0.13 \\
(ML, MO) & 0.86 & -0.24 & 0.06 & -0.56 \\
\hline
\end{tabular}
}
\caption{Rank correlations between LoBs within companies in terms of Spearman's $\rho$.}
\label{tab:RankCorrelationsBetweenLoBs}
\end{table}

The QQ plots in Figure~\ref{fig:NormalQQplots} indicate that the LoBs Home and Motor Other have normally distributed losses with mean close to zero. Let $U\tsub{H}^c$ and $U\tsub{MO}^c$ be the loss for company $c$ in LoB Home and Motor Other, respectively. We use the short forms: F for Folksam, I for If, LF for  L\"{a}nsf\"{o}rs\"{a}kringar and TH for Trygg-Hansa, and assume that the vector $(U\tsub{H}\tsup{F},U\tsub{H}\tsup{I},U\tsub{H}\tsup{LF},U\tsub{H}\tsup{TH},U\tsub{MO}\tsup{F},U\tsub{MO}\tsup{I},U\tsub{MO}\tsup{LF},U\tsub{MO}\tsup{TH})$ is drawn from a multivariate normal distribution with mean zero and covariance matrix
\[
\Sigma = \left( \begin{array}{cc}
\Sigma\tsub{H}       & \Sigma\tsub{H,MO}  \\
\Sigma\tsub{H,MO} & \Sigma\tsub{MO}     \\
\end{array} \right),
\]
where
\[
\Sigma\tsub{H} = \sigma\tsub{H}^2 \left( \begin{array}{cccc}
1 & \rho\tsub{H} & \rho\tsub{H} & \rho\tsub{H} \\
\rho\tsub{H} & 1 & \rho\tsub{H} & \rho\tsub{H} \\
\rho\tsub{H} & \rho\tsub{H} & 1 & \rho\tsub{H} \\
\rho\tsub{H} & \rho\tsub{H} & \rho\tsub{H} & 1 \\
\end{array} \right),
\]
\[
\Sigma\tsub{MO} = \sigma\tsub{MO}^2 \left( \begin{array}{cccc}
1 & \rho\tsub{MO} & \rho\tsub{MO} & \rho\tsub{MO} \\
\rho\tsub{MO} & 1 & \rho\tsub{MO} & \rho\tsub{MO} \\
\rho\tsub{MO} & \rho\tsub{MO} & 1 & \rho\tsub{MO} \\
\rho\tsub{MO} & \rho\tsub{MO} & \rho\tsub{MO} & 1 \\
\end{array} \right),
\]
and
\[
\Sigma\tsub{H,MO} = \sigma\tsub{H} \sigma\tsub{MO} \left( \begin{array}{cccc}
\rho_1 & \rho_2 & \rho_2 & \rho_2 \\
\rho_2 & \rho_1 & \rho_2 & \rho_2 \\
\rho_2 & \rho_2 & \rho_1 & \rho_2 \\
\rho_2 & \rho_2 & \rho_2 & \rho_1 \\
\end{array} \right).
\]
The assumptions leading to the matrices above are:
\begin{itemize}
\item for each LoB, the variance is the same for all companies, ie, $\Var(U\tsub{H}^c)=\sigma\tsub{H}^2$ and $\Var(U\tsub{MO}^c)=\sigma\tsub{MO}^2$ for all $c$,
\item for each LoB, the correlation between each pair of companies is the same, ie, $\Corr(U\tsub{H}^{c_1},U\tsub{H}^{c_2})=\rho\tsub{H}$ and $\Corr(U\tsub{MO}^{c_1},U\tsub{MO}^{c_2})=\rho\tsub{MO}$ for $c_1 \neq c_2$,
\item the correlation between Home and Motor Other is the same within all companies, ie, $\Corr(U\tsub{H}^c,U\tsub{MO}^c)=\rho_{1}$ for all $c$,
\item the correlation between Home in one company and Motor Other in another company is the same for each pair of companies, ie, $\Corr(U\tsub{H}^{c_1},U\tsub{MO}^{c_2})=\rho_{2}$ for $c_1 \neq c_2$.
\end{itemize}
Because of Folksam's outlier in the LoB Motor Other, we estimate the parameters both with and without the vector observed for accounting year 2001 in the sample. A likelihood ratio test strongly suggests that $\rho_1 = \rho_2$ in both cases, and the maximum likelihood parameter estimates are shown in Table~\ref{tab:HMOparameters}.

\begin{table}[tb]
\centering
\scalebox{\scaleFactorOne}{
\begin{tabular}{ld{1.3}d{1.3}d{1.2}d{1.2}d{1.2}}
\hline
Case & \multicolumn{1}{c}{$\hat{\sigma}\tsub{H}$} & \multicolumn{1}{c}{$\hat{\sigma}\tsub{MO}$} & \multicolumn{1}{c}{$\hat{\rho}\tsub{H}$} & \multicolumn{1}{c}{$\hat{\rho}\tsub{MO}$} & \multicolumn{1}{c}{$\hat{\rho}_1$}\\
\hline
Accounting year 2001 included      &  0.099 & 0.12  & 0.74 & 0.57 &  0.35 \\
Accounting year 2001 not included  &  0.10  & 0.096 & 0.75 & 0.52 &  0.64 \\
\hline
\end{tabular}
}
\caption{Maximum likelihood parameter estimates for Home and Motor Other.}
\label{tab:HMOparameters}
\end{table}

Denote the model for Home and Motor Other by $\M_1$, and let $\M_0$ be the reduction of $\M_1$ with $\rho_1 = 0$. The test statistic $D = 2 ( \ell(\M_1) - \ell(\M_0) )$, where $\ell(\M_i)$ is the maximum log-likelihood for model $\M_i$, is approximately $\chi^2_1$-distributed under the null hypothesis $\rho_1 = 0$ (see, eg, pp. 35--36 in \cite{Coles01}). If accounting year 2001 is included in the sample we observe $\hat{D}=2.73$, and since $\prob{D > 2.73} = 0.099$ under the null hypothesis, the choice between $\M_0$ and $\M_1$ is not obvious. In the SCR calculations in the following section we use $\M_1$ and the estimate $\hat{\rho}_{1} = 0.35$. This choice will yield higher capital requirements than $\M_0$, but due to the limited amount of data we choose to be conservative. If accounting year 2001 is not included in the sample we observe $\hat{D}=16.1$, so in this case the likelihood ratio test clearly suggests that $\rho_{1} \neq 0$.

For Illness and Accident we pool data from all companies, for Business Liability and Property we pool data from If, LF and Trygg-Hansa, and for Motor Liability we pool data from Folksam, If and LF. These decisions are based on the results of the Levene's tests conducted earlier in this section. The Motor Liability data is close to normally distributed. We assume zero mean and estimate the standard deviation $\sigma\tsub{ML}$ of the pooled data both with and without Folksam's observation from accounting year 2001. We get $\hat{\sigma}\tsub{ML} = 0.050$ if Folksam's 2001 observation is included, and $\hat{\sigma}\tsub{ML} = 0.025$ if it is not included.

\begin{figure}[tb]
\centering
\includegraphics[width=\textwidth]{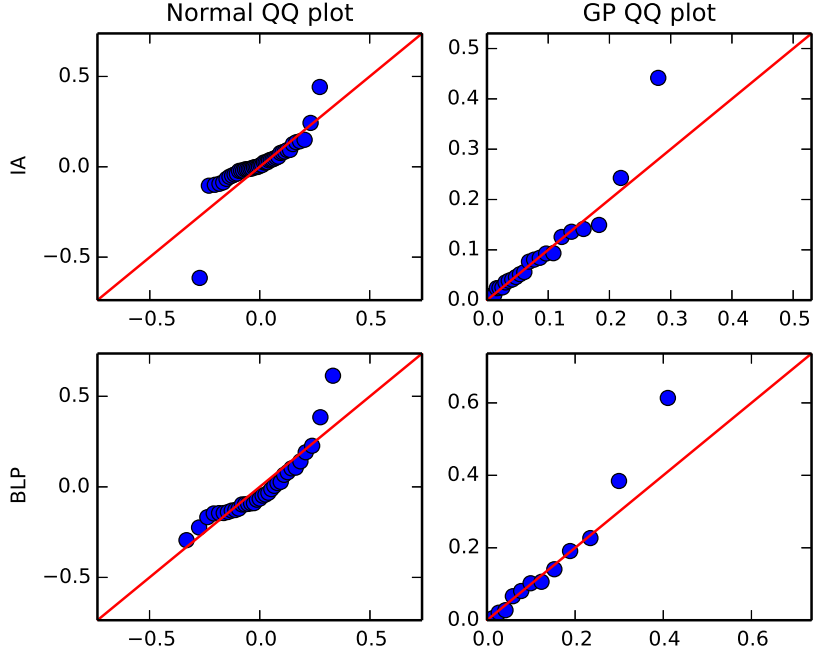}
\caption{Normal and generalized Pareto QQ plots for the pooled data of the LoBs Illness and Accident and Business Liability and Property.}
\label{fig:GenParetoQQplots}
\end{figure}

The yearly loss distributions for Illness and Accident and Business Liability and Property show signs of heavy-tailedness. For these LoBs we fit generalized Pareto (GP) distributions to the pooled positive losses using maximum likelihood. We get the shape and scale parameter estimates shown in Table~\ref{tab:GPparameters}, and the corresponding QQ plots are shown in Figure~\ref{fig:GenParetoQQplots}. It is worth emphasizing that it is not possible to justify the choice of the GP distribution by arguing that it is a limiting distribution. We choose the GP distribution since it has few parameters but is still flexible enough to capture different tail behaviors.

\begin{table}[tb]
\centering
\scalebox{\scaleFactorOne}{
\begin{tabular}{ld{2.2}d{1.3}}
\hline
Line of Business & \multicolumn{1}{c}{Shape ($\hat{\xi}$)} & \multicolumn{1}{c}{Scale ($\hat{\beta}$)} \\
\hline
Illness and Accident                               &  0     &  0.088 \\
Business Liability and Property (without Folksam)  &  0     &  0.16  \\
\hline
\end{tabular}
}
\caption{Maximum likelihood estimates of the generalized Pareto parameters.}
\label{tab:GPparameters}
\end{table}

\section{Modeling and SCR calculation} \label{sec:modeling}

In this section we set up simple internal models for each company. Moreover, we give a suggestion of how the marginal loss distribution of each LoB and the dependence between LoBs can be modeled using data from all four companies. We calculate SCR values for the four companies given these models.

For a given company, we denote today's liability predictions of incurred and not yet incurred claims for LoB $\ell$ by $R_0^{\ell}$ and $P_0^{\ell}$, respectively, and let $Y_0^{\ell}=R_0^{\ell} + P_0^{\ell}$. Suppose that we want to create a model for SCR calculation, but only have access to FSA data for our given company. Due to the limited amount of data we set up a very simple internal model where we assume that all LoBs are independent, and that the margins are normally distributed with zero mean. The 0.995 quantile of the total loss distribution, ie, the SCR for reserve and premium risk in Solvency II, in our internal model is given by
\[
\SCR\tsub{internal} = z_{0.995} \sqrt{ \sum_{\ell \in \{ \text{IA}, \text{H}, \text{BLP}, \text{ML}, \text{MO} \} } (Y_0^{\ell} s_{\ell})^2 },
\]
where $z_{0.995} \approx 2.58$ denotes the 0.995 quantile of the standard normal distribution and $s_{\ell}$ is the standard deviation of the normalized losses of LoB $\ell$. The SCR values calculated for this internal model using the liability predictions from the accounting year 2011 (see Table~\ref{tab:PremiumsAndPredictions}) and the sample standard deviations (see Table~\ref{tab:SampleStandardDeviations}) as parameters are shown in Table~\ref{tab:SCRSolvencyII}. Note that $V$ and $R_0$ are given by the actuaries in the reports, while $P_0$ depends on the author's loss ratio definition (see \eqref{eq:LossRatio} and \eqref{eq:P0}).

Now, we use the FSA data for all companies. Based on our findings in the previous sections, we assume that the total loss $X$ of our company is the sum of four independent losses
\[
X=X\tsub{IA} + X\tsub{BLP} + X\tsub{ML}+ X\tsub{H,MO},
\]
where $X\tsub{IA}$, $X\tsub{BLP}$ and $X\tsub{ML}$ are losses on the LoBs Illness and Accident, Business Liability and Property, and Motor Liability, respectively, and $X\tsub{H,MO}$ is the combined loss of the LoBs Home and Motor Other. Moreover, we assume that
\begin{itemize}
\item $X\tsub{H,MO}$ is normally distributed with mean zero and variance $(Y_0\tsup{H} \sigma\tsub{H})^2+ (Y_0\tsup{MO} \sigma\tsub{MO})^2 + 2 Y_0\tsup{H} Y_0\tsup{MO} \sigma\tsub{H} \sigma\tsub{MO} \rho_1$,
\item $X\tsub{ML}$ is normally distributed with mean zero and variance $(Y_0\tsup{ML} \sigma\tsub{ML})^2$,
\item for $\ell \in \{\text{IA},\text{BLP}\}$, $X_{\ell} = B_{\ell} Z_{\ell}$, where $B_{\ell}$ and $Z_{\ell}$ are independent, $Z_{\ell}$ is generalized Pareto distributed with shape parameter $\xi_{\ell}$ and scale parameter $Y_0^{\ell} \beta_{\ell}$, and $\prob{B_{\ell}=1} = \prob{B_{\ell}=-1} = 0.5$.
\end{itemize}
We use the liability predictions from the accounting year 2011 (see Table~\ref{tab:PremiumsAndPredictions}) together with the parameter assumptions shown in Table~\ref{tab:ParameterAssumptions} to calculate the 0.995 quantile of the distribution of the total loss $X$. The parameters in Model~1 and Model~2 are the estimates from the previous section when Folksam's outliers in Motor Liability and Motor Other are included and not included, respectively, in the sample. Notice that we use a different Motor Liability standard devation for Trygg-Hansa than for the other three companies. Levene's test showed us that Trygg-Hansa's variance was significantly different from the other companies' variance, so we use the sample standard deviation from Table~\ref{tab:SampleStandardDeviations} for Trygg-Hansa instead of the standard deviation of the pooled data sample. Also notice that we assume the same Business Liabiliy and Property parameters for all companies even though Folksam's data was considered different from the other companies' data and therefore not included in the pooled sample. However, since Folksam's BLP LoB is small compared to its other LoBs, this assumption does not affect the overall SCR of Folksam substantially.

\begin{table}[tb]
\centering
\scalebox{\scaleFactorOne}{
\begin{tabular}{cd{1.3}d{1.3}d{1.2}d{1.3}d{1.2}d{-1}d{1.3}d{-1}d{1.2}}
\hline
Model & \multicolumn{1}{c}{$\sigma\tsub{H}$} & \multicolumn{1}{c}{$\sigma\tsub{MO}$} & \multicolumn{1}{c}{$\rho_1$} & \multicolumn{1}{c}{$\sigma\tsub{ML}\tsup{F,I,LF}$} & \multicolumn{1}{c}{$\sigma\tsub{ML}\tsup{TH}$} & \multicolumn{1}{c}{$\xi\tsub{IA}$} & \multicolumn{1}{c}{$\beta\tsub{IA}$} & \multicolumn{1}{c}{$\xi\tsub{BLP}$} & \multicolumn{1}{c}{$\beta\tsub{BLP}$} \\
\hline
1  &  0.099 & 0.12  &  0.35 & 0.050  & 0.12 & 0 & 0.088 & 0 & 0.16 \\
2  &  0.10  & 0.096 &  0.64 & 0.025  & 0.12 & 0 & 0.088 & 0 & 0. 16\\
\hline
\end{tabular}
}
\caption{Parameter assumptions in the models based on pooled FSA data.}
\label{tab:ParameterAssumptions}
\end{table}

\begin{table}[tb]
\centering
\scalebox{\scaleFactorOne}{
\begin{tabular}{l|d{1.2}d{2.2}d{1.2}|d{1.2}d{2.2}d{1.2}|d{1.2}d{2.2}d{1.2}|d{1.2}d{2.2}d{1.2}}
\hline
 & \multicolumn{3}{c|}{Folksam} & \multicolumn{3}{c|}{If}&\multicolumn{3}{c|}{L\"{a}nsf\"{o}rs\"{a}kringar}& \multicolumn{3}{c}{Trygg-Hansa} \\
LoB &
 \multicolumn{1}{c}{$V$} & \multicolumn{1}{c}{$R_0$} & \multicolumn{1}{c|}{$P_0$} & \multicolumn{1}{c}{$V$} & \multicolumn{1}{c}{$R_0$} & \multicolumn{1}{c|}{$P_0$} & \multicolumn{1}{c}{$V$} & \multicolumn{1}{c}{$R_0$} & \multicolumn{1}{c|}{$P_0$} & \multicolumn{1}{c}{$V$} & \multicolumn{1}{c}{$R_0$} & \multicolumn{1}{c}{$P_0$} \\
\hline
IA & 1.49 & 5.05 & 1.11 & 0.64 & 1.07 & 0.42 & 1.30 & 3.18 & 1.08 & 2.53 & 6.00 & 1.51 \\
H & 2.67 & 1.12 & 1.76 & 1.63 & 0.59 & 1.10 & 3.51 & 1.61 & 2.51 & 1.49 & 0.65 & 1.08 \\
BLP & 0.26 & 0.14 & 0.18 & 1.85 & 2.27 & 1.02 & 5.13 & 3.71 & 3.22 & 1.67 & 1.26 & 1.05 \\
ML & 0.98 & 4.32 & 0.74 & 1.94 & 11.07 & 1.67 & 2.87 & 11.29 & 2.16 & 1.70 & 6.29 & 0.99 \\
MO & 1.96 & 0.18 & 1.13 & 3.50 & 0.34 & 2.19 & 3.62 & 0.60 & 2.45 & 2.11 & 0.39 & 1.44 \\
\hline
\end{tabular}
}
\caption{Earned premiums and initial predictions for accounting year 2011. All values in billion SEK.}
\label{tab:PremiumsAndPredictions}
\end{table}

For the LoBs Illness and Accident and Business Liability and Property, we assume distributions symmetric around zero where the absolute value is GP distributed with shape parameter $\xi=0$, ie, exponentially distributed. Thus, for $\ell \in \{\text{IA},\text{BLP}\}$, $\E[X_\ell] = 0$,
\[
\Var(X_\ell) = \E[X_\ell^2] = \E[Z_\ell^2] =  \Var(Z_\ell) + \E[Z_\ell]^2 = 2 \beta_\ell^2
\]
and
\[
F_{X_\ell}^{-1}(0.995) = F_{Z_\ell}^{-1}(0.99) = -\log(0.01) \beta_\ell \approx 3.26 \sqrt{\Var(X_\ell)},
\]
where $F^{-1}_{\cdot}$ denotes the inverse distribution function, ie, the quantile function. So, the ratio between the 0.995 quantile and the standard deviation is larger than for the normal distribution in these LoBs.


\section{The standard formula in Solvency II} \label{sec:standardFormula}

In this section we calculate SCR for the four companies using the standard formula in Solvency II (see \cite{Eiopa14} for details). We compare these values to the SCR values calculated using our models.

To calculate SCR for premium and reserve risk using the Solvency II standard formula, we must use the LoB segmentation defined in the directive. Illness and Accident includes two Solvency II LoBs; Medical Expense (ME) and Income Protection (IP) in the Health module. We assume that 25\% of the premiums (and reserves) correspond to Medical Expense and 75\% to Income Protection. Motor Liability and Motor Other correspond directly to the Solvency II LoBs Motor Vehicle Liability (MVL) and Other Motor (OM), respectively. Both these LoBs are in the Non-life module. We assume that 90\% of Home correspond to the Solvency II LoB Fire and Property Damage (FPD), and 10\% correspond to Third-Party Liability (TPL). For Business Liability and Property we assume that the proportions are 80\% and 20\%, respectively. Both Fire and Property Damage and Third-Party Liability are in the Non-life module. With these assumptions we get the Solvency II volume measures for premium and reserve risk, $V\tsub{prem}\tsup{LoB}$ and $V\tsub{res}\tsup{LoB}$, respectively, shown in Table~\ref{tab:SolvencyIIParameters}. In this table we also see the standard deviations specified by the regulators.

\begin{table}[tb]
\centering
\scalebox{\scaleFactorOne}{
\renewcommand{\arraystretch}{1.5}
\begin{tabular}{lccd{1.3}d{1.3}}
\hline
LoB (in Solvency II) & $V\tsub{prem}\tsup{LoB}$ & $V\tsub{res}\tsup{LoB}$ & \multicolumn{1}{c}{$\sigma\tsub{prem}\tsup{LoB}$} & \multicolumn{1}{c}{$\sigma\tsub{res}\tsup{LoB}$} \\
\hline
Medical Expense & $0.25 V\tsup{IA}$ & $0.25 R_0\tsup{IA}$ & 0.050 & 0.050 \\
Income Protection & $0.75 V\tsup{IA}$ & $0.75 R_0\tsup{IA}$ & 0.085 & 0.14 \\
Motor Vehicle Liability & $V\tsup{ML}$ & $R_0\tsup{ML}$ & 0.10 & 0.090 \\
Other Motor & $V\tsup{MO}$ & $R_0\tsup{MO}$ & 0.080 & 0.080 \\
Fire and Property Damage & $0.9 V\tsup{H} + 0.8 V\tsup{BLP}$ & $0.9 R_0\tsup{H} + 0.8 R_0\tsup{BLP}$ & 0.080 & 0.10 \\
Third-Party Liability  & $0.1 V\tsup{H} + 0.2 V\tsup{BLP}$ & $0.1 R_0\tsup{H} + 0.2 R_0\tsup{BLP}$ & 0.14 & 0.11 \\
\hline
\end{tabular}
}
\caption{Solvency II LoB segmentation and standard formula parameters.}
\label{tab:SolvencyIIParameters}
\end{table}

Using the values in Table~\ref{tab:SolvencyIIParameters} we calculate, for each LoB, the LoB volume measure
\begin{equation}
V\tsub{LoB} = V\tsub{prem}\tsup{LoB} + V\tsub{res}\tsup{LoB}
\label{eq:VLob}
\end{equation}
and the LoB standard deviation
\begin{equation}
\sigma\tsub{LoB} = \frac{\sqrt{(\sigma\tsub{prem}\tsup{LoB} V\tsub{prem}\tsup{LoB})^2 + 2 \alpha \sigma\tsub{prem}\tsup{LoB} \sigma\tsub{res}\tsup{LoB}  V\tsub{prem}\tsup{LoB} V\tsub{res}\tsup{LoB} + (\sigma\tsub{res}\tsup{LoB} V\tsub{res}\tsup{LoB})^2}}{V\tsub{LoB}},
\label{eq:sigmaLob}
\end{equation}
where the correlation coefficient $\alpha=0.5$ by assumption in the standard formula.

The LoBs Medical Expense and Income Protection belong to the sub-module NonSLT Health in the module Health. The volume measure for this sub-module is given by
\[
V\tsub{NonSLT Health} = V\tsub{ME} + V\tsub{IP},
\]
and the standard deviation is given by
\[
\sigma\tsub{NonSLT Health} = \frac{\sqrt{(\sigma\tsub{ME} V\tsub{ME})^2 + 2 \rho\tsub{ME,IP} \sigma\tsub{ME} \sigma\tsub{IP} V\tsub{ME} V\tsub{IP} + (\sigma\tsub{IP} V\tsub{IP})^2}}{V\tsub{NonSLT Health}},
\]
where the correlation coefficient $\rho\tsub{ME,IP}=0.5$ by assumption. Since we only consider premium and reserve risk, the SCR for the Health module is given by
\begin{equation}
\SCR\tsub{Health} = 3 \sigma\tsub{NonSLT Health} V\tsub{NonSLT Health}.
\label{eq:SCRHealth}
\end{equation}
The Solvency II assumption behind \eqref{eq:SCRHealth} is that the underlying risk follows a lognormal distribution. For the range of standard deviations considered in the Solvency II standard formula, the 0.995 quantile is approximately $3 \sigma$.

Now, define the set $S\tsub{Non-life}:=\{ \text{MVL}, \text{OM}, \text{FPD}, \text{TPL} \}$ containing the LoBs in the Non-life module. The volume measure for the premium and reserve risk of the Non-life module is given by
\[
V\tsub{Non-life} = \sum_{j \in S\tsub{Non-life}} V_{j},
\]
and the standard deviation is given by
\[
\sigma\tsub{Non-life} =\frac{1}{V\tsub{Non-life}} \sqrt{ \sum_{i \in S\tsub{Non-life}}  \sum_{j \in S\tsub{Non-life}} \rho_{i,j} \sigma_i \sigma_j V_i V_j },
\]
where the correlations are assumed to be $\rho\tsub{MVL,OM}=0.5$, $\rho\tsub{MVL,FPD}=0.25$, $\rho\tsub{MVL,TPL}=0.5$, $\rho\tsub{OM,FPD}=0.25$, $\rho\tsub{OM,TPL}=0.25$ and $\rho\tsub{FPD,TPL}=0.25$.

Since we only consider premium and reserve risk, the SCR for the Non-life module is given by
\[
\SCR\tsub{Non-life} = 3 \sigma\tsub{Non-life}  V\tsub{Non-life}.
\]
In the standard formula, the Health and Non-life modules are assumed to be independent. Hence, we get the overall SCR by
\[
\SCR = \sqrt{\SCR\tsub{Health}^2 + \SCR\tsub{Non-life}^2}.
\]
Again, note that we are only considering premium and reserve risk in this paper. The SCR values are calculated using the Solvency II standard formula are shown in Table~\ref{tab:SCRSolvencyII} together with the SCR values calculated using our model assumptions from the previous section.

\begin{table}[htb]
\centering
\scalebox{\scaleFactorOne}{
\renewcommand{\arraystretch}{1.5}
\begin{tabular}{ld{2.2}d{2.3}d{2.2}d{2.2}}
\hline
  & \multicolumn{1}{c}{Folksam} & \multicolumn{1}{c}{If} & \multicolumn{1}{c}{LF} & \multicolumn{1}{c}{Trygg-Hansa} \\
\hline
Predicted liabilities ($\sum_{\ell} Y_0^{\ell}$) & 15.73 & 21.74 & 31.81 & 20.66 \\
SCR(Internal model)   &  1.92  &  1.47  &  3.77  &  4.87  \\
SCR(Model 1)          &  2.69  &  2.99  &  5.63  &  3.93  \\
SCR(Model 2)          &  2.65  &  2.69  &  5.48  &  3.92  \\
SCR(Standard formula) &  2.84  &  4.54  &  6.02  &  3.73  \\
SCR(Internal model)/Predicted liabilities    & 0.12 & 0.068 & 0.12 & 0.23 \\
SCR(Model 1)/Predicted liabilities           & 0.17 & 0.14  & 0.18 & 0.19 \\
SCR(Model 2)/Predicted liabilities           & 0.17 & 0.12  & 0.17 & 0.19 \\
SCR(Standard formula)/Predicted liabilities  & 0.18 & 0.21  & 0.19 & 0.18 \\
\hline
\end{tabular}
}
\caption{SCR values calculated using our model assumptions and the Solvency II standard formula, respectively.}
\label{tab:SCRSolvencyII}
\end{table}

Now, for each Swedish LoB $\ell$, let $V^\ell$ be the sum of earned premiums over all companies and let $R_0^\ell$ be the sum of initial predictions of incurred claims over all companies for accounting year 2011 (see Table~\ref{tab:PremiumsAndPredictions}). Using these values together with the proportions and standard deviations in Table~\ref{tab:SolvencyIIParameters}, and equations \eqref{eq:VLob} and \eqref{eq:sigmaLob}, we get a benchmark value of the standard devation ($\sigma\tsub{LoB}$) for each Solvency II LoB.

The Swedish LoBs Motor Liability and Motor Other correspond directly to the Solvency II LoBs Motor Vehicle Liability and Other Motor, respectively. To get an idea about the standard formula assumptions regarding standard deviations for the Swedish LoBs Illness and Accident, Home and Business Liability and Property, we calculate
\[
\sigma_\ell\tsup{Standard formula} = \sqrt{(\sigma_{i_\ell} \pi_\ell)^2 
+ 2 \rho_{i_\ell,j_\ell} \sigma_{i_\ell} \sigma_{j_\ell} \pi_\ell (1 - \pi_\ell) + (\sigma_{j_\ell} (1 - \pi_\ell))^2},
\]
for $\ell \in \{ \text{IA}, \text{H}, \text{BLP} \}$, where $\pi_\ell$ and $(1 - \pi_\ell)$ are the proportions of the Swedish LoB $\ell$ allocated to the Solvency II LoBs $i_\ell$ and $j_\ell$, respectively, and $\rho_{i_\ell,j_\ell}$ is the correlation between the Solvency II LoBs $i_\ell$ and $j_\ell$ assumed in the standard formula. These standard deviations, and the corresponding 0.995 quantiles achieved by multiplying the standard deviation by $3$, are shown in Table~\ref{tab:StdevsAndQuantiles} together with the standard deviations and 0.995 quantiles of our models from the previous section.

\begin{table}[htb]
\centering
\scalebox{\scaleFactorOne}{
\renewcommand{\arraystretch}{1.5}
\begin{tabular}{l|d{1.3}d{1.2}d{1.3}|d{1.3}d{1.2}d{1.3}|d{1.3}d{-1}d{1.3}}
\hline
  & \multicolumn{3}{c|}{Model 1} & \multicolumn{3}{c|}{Model 2} & \multicolumn{3}{c}{Standard formula}   \\
    & \multicolumn{1}{c}{$\sigma$} & \multicolumn{1}{c}{$q_{0.995}/\sigma$} & \multicolumn{1}{c|}{$q_{0.995}$} & \multicolumn{1}{c}{$\sigma$} & \multicolumn{1}{c}{$q_{0.995}/\sigma$}  & \multicolumn{1}{c|}{$q_{0.995}$}  & \multicolumn{1}{c}{$\sigma$} & \multicolumn{1}{c}{$q_{0.995}/\sigma$} & \multicolumn{1}{c}{$q_{0.995}$}  \\
\hline
IA                 & 0.12   & 3.26 & 0.41  &  0.12   & 3.26 & 0.41   & 0.092  & 3 &  0.28 \\
H                  & 0.099  & 2.58 & 0.26  &  0.10   & 2.58 & 0.26   & 0.072  & 3 &  0.22 \\
BLP                & 0.23   & 3.26 & 0.74  &  0.23   & 3.26 & 0.74   & 0.070  & 3 &  0.21 \\
ML$\tsup{F,I,LF}$  & 0.050  & 2.58 & 0.13  &  0.025  & 2.58 & 0.064  & 0.084  & 3 &  0.25 \\
ML$\tsup{TH}$      & 0.12   & 2.58 & 0.31  &  0.12   & 2.58 & 0.31   & 0.084  & 3 &  0.25 \\
MO                 & 0.12   & 2.58 & 0.31  &  0.096  & 2.58 & 0.25   & 0.076  & 3 &  0.23 \\
\hline
\end{tabular}
}
\caption{Assumptions about standard deviations ($\sigma$) and 0.995 quantiles ($q_{0.995}$) in our models and the Solvency II standard formula, respectively.}
\label{tab:StdevsAndQuantiles}
\end{table}

\section{Discussion} \label{sec:discussion}

One major finding in the analysis is that dependence between LoBs seems to be weaker than what is assumed in the Solvency II standard formula. The only signs of dependence we see in the FSA data is between the LoBs Home and Motor Other, and between companies within these LoBs. These two light-tailed LoBs typically have many small claims with short time between accident and final payment. More research is needed in order to find out what causes the dependence between companies within these LoBs, but one hypothesis is that competition has caused premiums to converge, and since the companies to a large extent are exposed to the same risks in these LoBs the relative losses also converge. Losses on the LoBs Illness and Accident and Motor Liability are mainly due to revaluation of old claims. To get dependence between these LoBs, or between companies within any of these LoBs, there should be some external factor (for example a legal change) that causes all revaluations to go in the same direction. However, we do not see any such sign in the FSA data.

The SCR values calculated using Model~1 and Model~2 of this paper are in the neighborhood of the values calculated using the Solvency II standard formula for Folksam, LF and Trygg-Hansa (see Table~\ref{tab:SCRSolvencyII}). For If, however, the SCR is substantially lower in our models than in the standard formula. Due to its large motor portfolio, the total loss of If is extremely dependent on the Motor Liability distribution which is light-tailed with low variance in our models. 
 
The standard formula assumes positive dependencies between LoBs that are hard to motivate from what we see in the FSA data. However, the SCR values calculated using the standard formula are not too far away from the values calculated using our models, so the marginal distributions in the standard formula must have either lower variance or lighter tails than what we assume in our models. In Table~\ref{tab:StdevsAndQuantiles} we see that the standard formula assumes lower variance in all LoBs except Motor Liability, lighter tails in Illness and Accident and Business Liability and Property, but heavier tails in Home, Motor Liability and Motor Other. Since dependence structures are very difficult to estimate from a small data sample, my suggestion is to assume independence when no clear dependence is seen in data, and use conservative estimates for the marginal distributions to (at least to some extent) offset the dependence uncertainty. More research, and data, is needed in order to determine the marginal one-year loss distributions as well as dependence structures with less uncertainty.
 
An interesting implication of using data from several companies to create a standard model for SCR calculation (for example the Solvency II standard formula) is that high prediction uncertainty in one LoB for one company may imply higher SCR values for all companies. To get rid of the implicit dependence of actuaries and/or executives in other companies that follows from using the standard model, companies should be encouraged to develop their own internal models for SCR calculation. A company that has good models for liability predictions should be able to motivate the use of an internal model that yields a lower SCR value than the standard model. However, a company that adjusts liability predictions to smooth gains and losses over time (and thereby increases the liability prediction variance) may be better off using the standard model.

All liability predictions in this paper are done without discounting. Extending the FSA reports to include full run-off triangles would make discounting possible. Moreover, with full triangles it would be possible to investigate the dependencies between the liability and the asset side of an insurer's balance sheet. The dependence between insurance losses and the interest rates used for discounting would be of particular interest to investigate in a future analysis.

\section*{Acknowledgments}

The author thanks Erik Elvers at the Swedish FSA for suggesting to analyze the FSA data, for explaining how to interpret the reports and for giving suggestions on how to translate the Swedish insurance classes into the Solvency II segmentation. The author also thanks: Filip Lindskog, Holger Rootz\'{e}n, Gunnar Andersson, Bengt von Bahr and {\AA}sa Larson. The author acknowledges financial support from the Swedish Research Council (Reg. No. 2009-5523).

\end{document}